\documentclass[runningheads]{llncs}
\usepackage{graphicx}
\usepackage{amsmath}
\usepackage{multirow}
\usepackage[table,xcdraw]{xcolor}
\usepackage{comment}
\usepackage[normalem]{ulem}
\useunder{\uline}{\ul}{}
\usepackage[misc]{ifsym}
\usepackage{orcidlink}

\begin{document}
\title{Deep Learning Models to Automate the Scoring of Hand Radiographs for Rheumatoid Arthritis}
\titlerunning{Deep Learning Models to Score Hand Radiographs for Rheumatoid Arthritis}

\author{Zhiyan Bo\inst{1}\textsuperscript{(\Letter)}\orcidlink{0009-0002-6458-3156}\and
Laura C. Coates\inst{2}\orcidlink{0000-0002-4756-663X}
\and Bart\l omiej W. Papie\.z\inst{1}\orcidlink{0000-0002-8432-2511}
}
\authorrunning{Z. Bo et al.}

\institute{Big Data Institute, Nuffield Department of Population Health, University of Oxford, Oxford, UK \\
\email{zhiyan.bo@reuben.ox.ac.uk} \&
\email{bartlomiej.papiez@bdi.ox.ac.uk}\\
\and
Nuffield Department of Orthopaedics, Rheumatology and Musculoskeletal Sciences, University of Oxford, Oxford, UK}
\maketitle            

\begin{abstract}
The van der Heijde modification of the Sharp (SvdH) score is a widely used radiographic scoring method to quantify damage in Rheumatoid Arthritis (RA) in clinical trials. However, its complexity with a necessity to score each individual joint, and the expertise required limit its application in clinical practice, especially in disease progression measurement. 
In this work, we addressed this limitation by developing a bespoke, automated pipeline that is capable of predicting the SvdH score and RA severity from hand radiographs without the need to localise the joints first.
Using hand radiographs from RA and suspected RA patients, we first investigated the performance of the state-of-the-art architectures in predicting the total SvdH score for hands and wrists and its corresponding severity class.
Secondly, we leveraged publicly available data sets to perform transfer learning with different finetuning schemes and ensemble learning, which resulted in substantial improvement in model performance being on par with an experienced human reader.
The best model for RA scoring achieved a Pearson's correlation coefficient (PCC) of 0.925 and root mean squared error (RMSE) of 18.02, while the best model for RA severity classification achieved an accuracy of 0.358 and PCC of 0.859. Our score prediction model attained almost comparable accuracy with experienced radiologists (PCC = 0.97, RMSE = 18.75). Finally, using Grad-CAM, we showed that our models could focus on the anatomical structures in hands and wrists which clinicians deemed as relevant to RA progression in the majority of cases.

\end{abstract}

\section{Introduction}
With a prevalence of about 1\%, more than 400,000 people in the UK are diagnosed with Rheumatoid Arthritis (RA), an autoimmune disease that causes joint inflammation which could progress to structural damage, pain and disability in patients \cite{NICE}. Its classic presentation is symmetric arthritis involving multiple joints in both hands, feet and wrists with bone erosions (indicated by blue arrows in Fig.~\ref{fig1}(A)) and joint space narrowing (JSN) (indicated by blue boxes in Fig.~\ref{fig1}(A)) \cite{BMJ}. As a widely available tool in clinical settings, plain radiography is the gold standard for RA progression evaluation and is commonly used in its diagnosis, classification and monitoring \cite{BMJ,Welsing2004}.

Several diagnostic and RA quantification criteria have been proposed, with the most commonly used criteria for radiographic scoring being the van der Heijde modification of the Sharp (SvdH) score, which is a standard damage quantification method in clinical trials due to its good intra- and inter-observer reliability, and sensitivity to changes~\cite{Landewe2016,Radu2021,Salaffi2019,VanDerHeijde1999}.
It includes two scores for bone erosions and JSN in hands, wrists and forefeet. Due to SvdH’s complexity, it takes an experienced radiologist or rheumatologist around 4 minutes to score single-hand radiography; moreover imaging artefacts and superimposition can lead to interobserver disagreements especially when measuring small changes \cite{Boini2001,VanDerHeijde1999_2,VanDerHeijde1999}.
The above shortcomings limit the method’s clinical application and hinder its effectiveness in clinical trials. 
In routine clinical practice, rheumatologists simply compare images instead of generating and comparing the scores.

Multiple Deep Learning (DL)-based approaches have been developed for RA scoring using ultrasound and X-ray \cite{Bird2022,Gilvaz2023,Imtiaz2022}. Most of those methodologies consist of a joint detection or segmentation stage followed by a severity score prediction stage \cite{Hemalatha2019,Ho2018,Tan2021}. For example, a pipeline proposed adopts a contouring technique to segment the synovial regions from ultrasound images and utilises a convolutional neural network (CNN) to classify the synovial regions into different severity grades \cite{Hemalatha2019}. 
An automatic joint detection (using You Only Look Once v4) and SvdH-based classification system (using EfficientNet-B1) was developed for hand X-ray images \cite{WangSu2022}. However, these methods assign a severity class to a joint instead of a score, which reduces the resolution of the prediction.

DL-based frameworks that directly predict the SvdH score have been explored as well \cite{Bird2022,Parashar2023}. Hirano et al. focused on the finger joints and developed a pipeline that first detects them by a cascade classifier using Haar-like features and then scores each joint for erosion and JSN by a CNN \cite{Hirano2019}. In the RA-2–Dialogue for Reverse Engineering Assessment and Methods (RA2-DREAM Challenge), CNNs predicting the overall SvdH score from hands and feet X-rays as well as the joint-by-joint and overall JSN and erosion scores separately were investigated \cite{Sun2022}. Among the published solutions, a multi-task DL model adopting U-Net and EfficientNet-B5 architecture that localises the joints and predicts the SvdH scores for JSN and erosion was proposed \cite{Maziarz2022}. 
Similarly, another proposed pipeline detects the joints using U-Net-based heatmap regression, crops the joint image, and then predicts the amount of damage by EfficientNet-B0~\cite{Honda2023}.

Although the above methods have demonstrated promising performance, their suitability for clinical applications is still uncertain as it could be challenging to localise individual joints accurately. 
Also, late-stage RA patients may find it painful to straighten their fingers with deformities during an X-ray and algorithms might have difficulty delineating severely damaged bones and joints.
Therefore methodologies to predict overall scores might be practically useful and ease the procedure of disease severity quantification. Instead of focusing on each joint, a model ResNet-Dwise50 combining ResNet-50 and MobileNetV2 which predicts an overall SvdH score for hand X-ray was developed \cite{Wang2022}. They concluded that using depth-wise separable convolutions and replacing the residual blocks (RBs) with inverted residual blocks (IRBs) resulted in the observed performance improvement compared to baseline ResNet-50.

In this work, we focused on using whole bilateral posteroanterior hand radiographs as input for RA damage quantification and proposed two ensemble models that combine three CNNs to predict 1) the SvdH score and 2) the corresponding RA severity class.
Our contributions can be summarised as follows.
We first investigated the performance of the state-of-the-art CNNs: MobileNetV2 \cite{Sandler2018} and ResNet \cite{He2016} in a publicly available RA hand X-ray dataset. Secondly, we meticulously developed efficient approaches to improve the performance of our models when compared with the state-of-the-art method \cite{Wang2022} using transfer learning.
Then, ensemble technique was applied, further improving the accuracy of RA scoring to be on par with human readers.
Finally, we addressed the limited interpretability of the DL methods using entire radiographs as input by performing Gradient-weighted Class Activation Mapping (Grad-CAM) \cite{Selvaraju2017}. The heatmaps generated were qualitatively assessed to evaluate the clinical potential of our models.

\section{Data and Methods}
\subsection{Data}
This project utilised a dataset which includes 3,818 hand X-ray images (see exemplar image in Fig.~\ref{fig1}(A)) collected from RA patients or suspected RA patients (see details \cite{Wang2022}). The average of the SvdH scores reported by two experienced radiologists was provided, alongside patients’ age and gender. Most of the samples were older than 30 and 82.6\% of them were females. The data publisher also evaluated the inter-rater differences, which demonstrated a Pearson’s correlation coefficient (PCC; Eq.\ref{eq_PCC}) of 0.97, mean absolute error (MAE; Eq.\ref{eq_MAE}) of 12.24, and root mean squared error (RMSE; Eq.\ref{eq_RMSE}) of 18.75. The dataset was split into training, validation and test sets, each including 2700, 760 and 358 images. 

\subsection{Pre-processing}
\begin{figure}[!ht]
\centering
\includegraphics[width=0.9\textwidth]{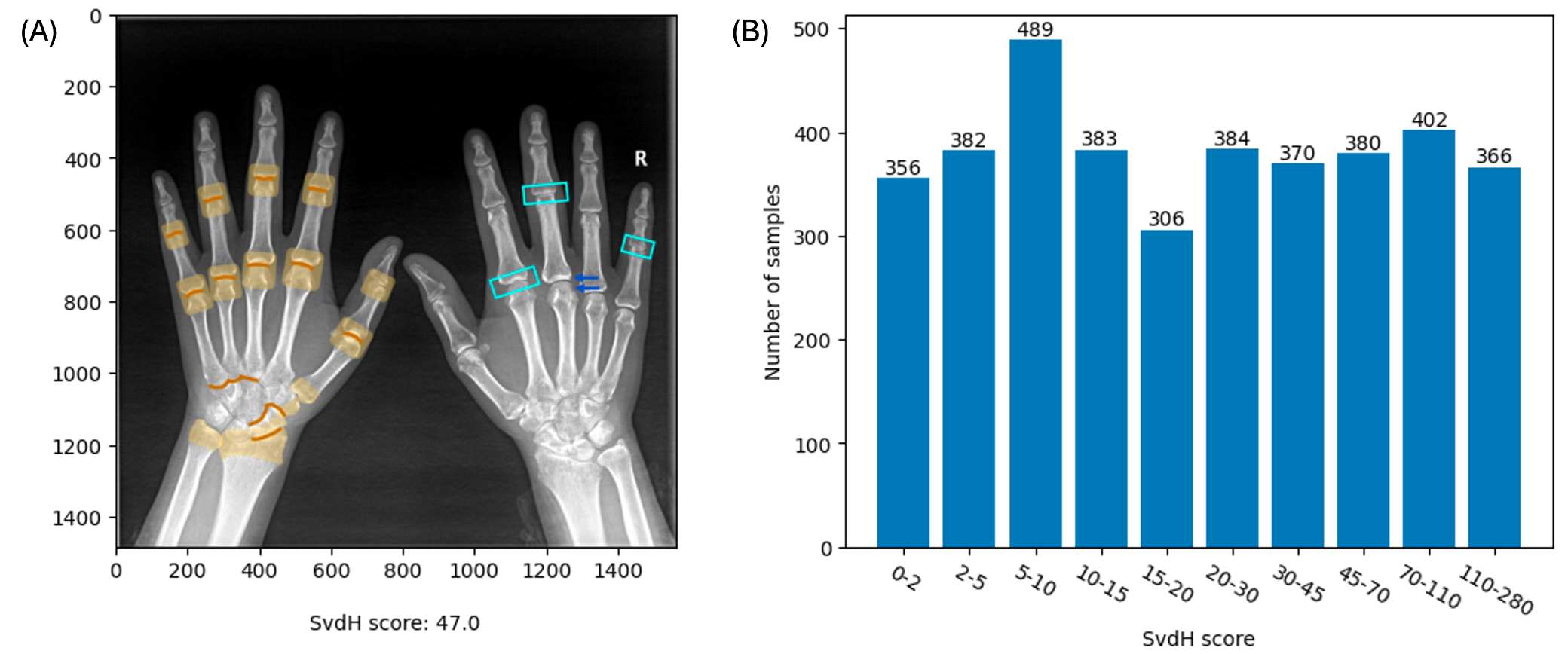}
\caption{(A) An example of the radiographs from Wang et al. \cite{Wang2022} with an average SvdH score of 47. The bones and joints assessed in the left hand and wrist are highlighted in yellow and orange for reference. Examples of bone erosions and JSN are indicated by blue arrows and boxes. (B) The SvdH class distribution of samples used in this paper.} \label{fig1}
\end{figure}

The SvdH score looks at 16 areas (yellow in Fig.~\ref{fig1}(A)) and 15 joints (orange in Fig.~\ref{fig1}(A)) in each hand and wrist for erosion and JSN assessment. The erosion score ranges from 0 to 5 while the JSN score ranges from 0 to 4, so the total score accounting for hands and wrists ranges from 0 to 280 though scores $>$ 140 are less commonly seen \cite{VanDerHeijde1999}. More details are provided in the supplementary material. For direct prediction, the scores of the images were standardised via the z-score method to have a mean of 0 and a standard deviation (SD) of 1. 
In addition to score prediction, we investigated damage quantification as a classification problem. As in clinical settings, a large proportion of cases were diagnosed or maintained at relatively early stages, the scores were split into 10 classes (Fig.~\ref{fig1}(B)), with smaller ranges at the lower end of the score range.

To increase the sample size and diversity, the training images were first resized to $1024 \times 1024$ and augmented with random horizontal flip, random intensity changes with a scale in the range of 0.9 - 1.1, and random rotation by one of the given angles [0\textdegree, 90\textdegree, 180\textdegree, 270\textdegree] to avoid exclusion of assessed anatomical structures. Finally image pixel intensities were normalised with the mean and SD of the training set. For the validation and test sets, only image resizing and normalisation were performed.

\subsection{Models}
In \cite{Wang2022} ResNet-50 and MobileNetV2 were used as the baseline models, thus for comparison, the two models were selected for our study. 
The ResNet architecture adopts skip connections that enable a model layer to learn residual functions regarding the layer inputs to address the issue of vanishing gradient \cite{He2016}. To investigate the effect of model complexity on performance, ResNet-34, a version of ResNet with fewer layers, was also used. 
MobileNetV2 uses IRBs with depthwise separable convolution and linear bottleneck. With fewer model parameters, it demonstrated good efficiency and performance in many image analysis tasks \cite{Sandler2018}.
In this work separate sets of models were trained for the regression and classification tasks, with the output size of the models' last fully connected layer changed to one for regression and 10 for classification.  

\subsubsection{Loss function.}
For training classifiers, the cross-entropy loss was employed, while two distinct loss functions were chosen for training the regression models.
The first loss was the mean squared error (MSE) loss. The second one was the smooth loss (SL) proposed in \cite{Wang2022}, which combines MSE and MAE loss defined as follows.

\begin{equation}
SL\ =\ \sqrt{\frac{1}{n}\sum_{i=1}^{n}{smooth\left(y_i-{\hat{y}}_i\right)}} \end{equation}  
in which \textit{smooth} is defined as:
\begin{equation}
smooth\left(x\right) =\left\{\begin{matrix}ax^2 &if\ \left|x\right|<c\\ 
\left|x\right|-b&otherwise\\\end{matrix}\right.\end{equation}
where $a$=0.6, $b$=0, and $c$=1.0. These parameters gave the best performance in \cite{Wang2022}, so we adopted the same values. 
In exploratory testing, we obtained more accurate models with MSE loss so it was used for further model training and transfer learning.

\subsection{Transfer learning}
Transfer learning has been widely adopted in medical imaging to improve the performance of models \cite{Salehi2023}. In this work, transfer learning was performed using the RSNA Pediatric Bone Age Challenge dataset \cite{Halabi2019}, which includes 14,036 single-hand X-ray images with skeletal age in months. 12,611 (90\%) images were used for training and the remaining 1,425 (10\%) for validation. Z-score standardisation of bone age, image augmentation, MSE loss and a batch size of 4 were adopted.
ResNets and MobileNetV2 were trained for 50 epochs for bone age prediction, and then the weights were transferred for SvdH scoring and RA severity classification. Two types of model-tuning methods were experimented with:
1)	Finetuning, where all model layers are trainable;
2)	Freezing the first few layers during training and only tuning the higher-level parameters.

Two layer-freezing schemes were picked for each model to find a balance between training accuracy and generalisability. For ResNets, we tuned a set of models while freezing layers up to and including the first stack of residual blocks (RBs; RBs-1) and a second set while freezing layers up to and including the second stack of RBs (RBs-2). For MobileNetV2, two models were tuned separately while freezing up to and including the second stack of inverted residual blocks (IRBs; IRBs-2) and while freezing up to and including the third stack of IRBs (IRBs-3). With RBs-1 and IRBs-2, we investigated the effect of preserving the low-level features that could be common for both the new and the original tasks as similar types of images were used. Bone age assessment and SvdH scoring look at some similar anatomical features, such as joint space, so we also experimented with preserving the middle-level features as well with RBs-2 and IRBs-3.    

\subsection{Ensemble learning}
As a widely adopted ensemble technique, stacking was shown to bring substantial performance gains in medical image classification tasks \cite{Muller2022}. It assembles the outputs of independent models and feeds them into an extra algorithm for prediction. In this work, we applied a linear transformation to the outputs from the best-performing ResNet-50, ResNet-34 and MobileNetV2 to predict the SvdH score or probability of each severity class. As classifiers each produce 10 outputs for 10 classes, two stacking methods were attempted: 1) All classes: using all outputs of three models to make predictions for 10 classes to account for inter-class associations; 2) Single class: using class-specific outputs only for prediction.

\subsection{Model explainability}
Since the SvdH scoring scheme looks at specific joints and bones in the hand X-ray, we visually assessed the activation maps, which infer the contribution of pixels in an image to the final prediction. Examples of true positive (TP), true negative (TN), false positive (FP) and false negative (FN) were selected. For regression, TN and TP were defined as healthy (SvdH $<$ 5) and severe (SvdH $>$ 200) cases with an absolute prediction error smaller than 10. FP and FN were defined as cases with an absolute prediction error greater than 50. Grad-CAM was applied to the activation map computed before the fully connected layer. Finally, we invited an experienced rheumatologist to comment on whether the highlighted anatomical structures demonstrate representative pathology. 

\subsection{Evaluation}
For regression, models were evaluated in the test set by measuring the degree of agreement between the predicted scores and the averaged true scores. Metrics used were PCC, MAE and RMSE defined as follows.
\begin{equation} \label{eq_PCC}
    PCC\ =\ \frac{cov\left(X,Y\right)}{\sqrt{var\left(X\right)}\sqrt{var\left(Y\right)}}      
\end{equation}
\begin{equation}\label{eq_MAE}
MAE\ =\ \frac{1}{n}\sum_{i=1}^{n}\left|x_i-y_i\right|
\end{equation}
\begin{equation} \label{eq_RMSE}
RMSE\ =\sqrt{\frac{1}{n}\sum_{i=1}^{n}\left(x_i-y_i\right)^2}
\end{equation}
For classification, models were evaluated using confusion matrix accuracy, balanced accuracy (BA), and PCC, MAE and RMSE to account for the ordinal relationships between classes. The performance of the best regression model in severity classification was evaluated by assigning samples to different predicted classes according to their predicted scores.

\subsection{Implementation details}
Stochastic gradient descent optimiser with an initial learning rate of 0.001, weight decay of 0.001, and momentum of 0.9 was used. Models with the best validation performance during training were selected. All regression models were trained on an NVIDIA Tesla V100 SXM2 GPU with 32GB of memory to compare their training time. For loss function comparison, models were trained for 100 epochs with a batch size of 4. With MSE loss, we then investigated the effect of using a larger batch size (16). For transfer learning, models were trained with a batch size of 4.
Classification models were trained on different NVIDIA GPUs as we did not plan to measure the training time. All baseline models were trained for 100 epochs with a batch size of 4 or 16. For transfer learning, batch size = 4 was adopted.

\section{Results}

\subsection{Regression task: SvdH score prediction}
\subsubsection{Baseline.}
Table~\ref{tab1} displays the results for baseline regression models without pretraining. MSE loss achieved better performances in terms of all three metrics, with fewer fluctuations in the models’ validation performance during training. The average time taken by one epoch was 11:28 min, 12:28 min and 13:45 min for MobileNetV2, ResNet-34 and ResNet-50 when trained with batch size = 4. Increasing the batch size to 16 did not substantially change the performance of ResNet-34 and ResNet-50. Their PCC increased while their MAE and RMSE deteriorated. On the other hand, improvements in two of the three metrics were observed for MobileNetV2. Its PCC increased from 0.877 to 0.903, RMSE decreased from 22.72 to 20.48, and MAE increased slightly from 14.84 to 14.99. Among all models without transfer learning, the MobileNetV2 trained with a batch size of 16 achieved the best performance. However, using a larger batch size increased the speed of convergence but worsened the stability of model performance during training, especially after reaching the plateau in validation loss. For result consistency and model training stability, batch size = 4 was adopted for transfer learning of all three models presented below. 

\begin{table}[!ht]
\centering
\caption{Model performance in SvdH score prediction. The top two values of a metric are made bold and underlined. Among baseline models, MobileNetV2 demonstrated the highest accuracy. With transfer learning, ResNet-50:RBs-1 achieved the best performance among all independent models. Ensemble learning yielded further performance improvement and the best regression model was obtained with batch size = 16.}\label{tab1}
\resizebox{0.72\textwidth}{!}{
\begin{tabular}{llllll}
\multicolumn{6}{l}{\cellcolor[HTML]{F8CCA6}\textbf{Baseline}}\\ \hline
\multicolumn{1}{|l|}{Model}                                                & \multicolumn{1}{l|}{Loss function}                                & \multicolumn{1}{l|}{Batch size} & \multicolumn{1}{l|}{PCC}            & \multicolumn{1}{l|}{MAE}            & \multicolumn{1}{l|}{RMSE}           \\ \hline
\multicolumn{1}{|l|}{}                                                     & \multicolumn{1}{l|}{Smooth}                                       & \multicolumn{1}{l|}{4}          & \multicolumn{1}{l|}{0.838}          & \multicolumn{1}{l|}{16.87}          & \multicolumn{1}{l|}{26.02}          \\ \cline{2-6} 
\multicolumn{1}{|l|}{}                                                     & \multicolumn{1}{l|}{}                                             & \multicolumn{1}{l|}{4}          & \multicolumn{1}{l|}{0.885}          & \multicolumn{1}{l|}{15.33}          & \multicolumn{1}{l|}{22.72}          \\ \cline{3-6} 
\multicolumn{1}{|l|}{\multirow{-3}{*}{ResNet-50}}                          & \multicolumn{1}{l|}{\multirow{-2}{*}{MSE}}                        & \multicolumn{1}{l|}{16}         & \multicolumn{1}{l|}{{\ul{0.895}}}    & \multicolumn{1}{l|}{17.65}          & \multicolumn{1}{l|}{23.26}          \\ \hline
\multicolumn{1}{|l|}{}                                                     & \multicolumn{1}{l|}{Smooth}                                       & \multicolumn{1}{l|}{4}          & \multicolumn{1}{l|}{0.828}          & \multicolumn{1}{l|}{16.9}           & \multicolumn{1}{l|}{27.14}          \\ \cline{2-6} 
\multicolumn{1}{|l|}{}                                                     & \multicolumn{1}{l|}{}                                             & \multicolumn{1}{l|}{4}          & \multicolumn{1}{l|}{0.876}          & \multicolumn{1}{l|}{15.11}          & \multicolumn{1}{l|}{{\ul{22.7}}}     \\ \cline{3-6} 
\multicolumn{1}{|l|}{\multirow{-3}{*}{ResNet-34}}                          & \multicolumn{1}{l|}{\multirow{-2}{*}{MSE}}                        & \multicolumn{1}{l|}{16}         & \multicolumn{1}{l|}{0.885}          & \multicolumn{1}{l|}{17.1}           & \multicolumn{1}{l|}{22.89}          \\ \hline
\multicolumn{1}{|l|}{}                                                     & \multicolumn{1}{l|}{Smooth}                                       & \multicolumn{1}{l|}{4}          & \multicolumn{1}{l|}{0.877}          & \multicolumn{1}{l|}{{\ul{14.96}}}    & \multicolumn{1}{l|}{23.11}          \\ \cline{2-6} 
\multicolumn{1}{|l|}{}                                                     & \multicolumn{1}{l|}{}                                             & \multicolumn{1}{l|}{4}          & \multicolumn{1}{l|}{0.877}          & \multicolumn{1}{l|}{\textbf{14.84}} & \multicolumn{1}{l|}{22.72}          \\ \cline{3-6} 
\multicolumn{1}{|l|}{\multirow{-3}{*}{MobileNetV2}}                        & \multicolumn{1}{l|}{\multirow{-2}{*}{MSE}}                        & \multicolumn{1}{l|}{16}         & \multicolumn{1}{l|}{\textbf{0.903}} & \multicolumn{1}{l|}{14.99}          & \multicolumn{1}{l|}{\textbf{20.48}} \\ \hline
\multicolumn{6}{l}{\cellcolor[HTML]{B5EAF9}\textbf{Transfer learning (finetuned with MSE loss)}} \\ \hline
\multicolumn{1}{|l|}{Model}                                                & \multicolumn{1}{l|}{Layer freezing}                               & \multicolumn{1}{l|}{Batch size} & \multicolumn{1}{l|}{PCC}            & \multicolumn{1}{l|}{MAE}            & \multicolumn{1}{l|}{RMSE}           \\ \hline
\multicolumn{1}{|l|}{}                                                     & \multicolumn{1}{l|}{No}                                           & \multicolumn{1}{l|}{4}          & \multicolumn{1}{l|}{0.907}          & \multicolumn{1}{l|}{\textbf{12.97}} & \multicolumn{1}{l|}{20.14}          \\ \cline{2-6} 
\multicolumn{1}{|l|}{}                                                     & \multicolumn{1}{l|}{RBs-1}                                        & \multicolumn{1}{l|}{4}          & \multicolumn{1}{l|}{\textbf{0.918}} & \multicolumn{1}{l|}{13.58}          & \multicolumn{1}{l|}{\textbf{18.88}} \\ \cline{2-6} 
\multicolumn{1}{|l|}{\multirow{-3}{*}{ResNet-50}}                          & \multicolumn{1}{l|}{RBs-2}                                        & \multicolumn{1}{l|}{4}          & \multicolumn{1}{l|}{0.908}          & \multicolumn{1}{l|}{{\ul{13.53}}}    & \multicolumn{1}{l|}{19.86}          \\ \hline
\multicolumn{1}{|l|}{}                                                     & \multicolumn{1}{l|}{No}                                           & \multicolumn{1}{l|}{4}          & \multicolumn{1}{l|}{{\ul{0.912}}}    & \multicolumn{1}{l|}{13.84}          & \multicolumn{1}{l|}{{\ul{19.78}}}    \\ \cline{2-6} 
\multicolumn{1}{|l|}{}                                                     & \multicolumn{1}{l|}{RBs-1}                                        & \multicolumn{1}{l|}{4}          & \multicolumn{1}{l|}{0.902}          & \multicolumn{1}{l|}{14.74}          & \multicolumn{1}{l|}{20.57}          \\ \cline{2-6} 
\multicolumn{1}{|l|}{\multirow{-3}{*}{ResNet-34}}                          & \multicolumn{1}{l|}{RBs-2}                                        & \multicolumn{1}{l|}{4}          & \multicolumn{1}{l|}{0.9}            & \multicolumn{1}{l|}{13.68}          & \multicolumn{1}{l|}{20.72}          \\ \hline
\multicolumn{1}{|l|}{}                                                     & \multicolumn{1}{l|}{No}                                           & \multicolumn{1}{l|}{4}          & \multicolumn{1}{l|}{0.898}          & \multicolumn{1}{l|}{14.02}          & \multicolumn{1}{l|}{20.95}          \\ \cline{2-6} 
\multicolumn{1}{|l|}{}                                                     & \multicolumn{1}{l|}{IRBs-2}                                       & \multicolumn{1}{l|}{4}          & \multicolumn{1}{l|}{0.898}          & \multicolumn{1}{l|}{13.97}          & \multicolumn{1}{l|}{20.81}          \\ \cline{2-6} 
\multicolumn{1}{|l|}{\multirow{-3}{*}{MobileNetV2}}                        & \multicolumn{1}{l|}{IRBs-3}                                       & \multicolumn{1}{l|}{4}          & \multicolumn{1}{l|}{0.907}          & \multicolumn{1}{l|}{13.59}          & \multicolumn{1}{l|}{19.93}          \\ \hline
\multicolumn{6}{l}{\cellcolor[HTML]{BFFF99}\textbf{Ensemble learning (finetuned with MSE loss)}} \\ \hline
\multicolumn{2}{|l|}{Model combination}                                                                                                        & \multicolumn{1}{l|}{Batch size} & \multicolumn{1}{l|}{PCC}            & \multicolumn{1}{l|}{MAE}            & \multicolumn{1}{l|}{RMSE}           \\ \hline
\multicolumn{2}{|l|}{}                                                                                                                         & \multicolumn{1}{l|}{4}          & \multicolumn{1}{l|}{{\ul{0.925}}}    & \multicolumn{1}{l|}{{\ul{12.63}}}    & \multicolumn{1}{l|}{{\ul{18.17}}}    \\ \cline{3-6} 
\multicolumn{2}{|l|}{\multirow{-2}{*}{\begin{tabular}[c]{@{}l@{}}ResNet-50:RBs-1 \& ResNet-34:finetuned \&\\ MobileNetV2:IRBs-3\end{tabular}}} & \multicolumn{1}{l|}{16}         & \multicolumn{1}{l|}{\textbf{0.925}} & \multicolumn{1}{l|}{\textbf{12.57}} & \multicolumn{1}{l|}{\textbf{18.02}} \\ \hline
\end{tabular}
}
\end{table}

\subsubsection{Transfer learning.}
\begin{figure}[!ht]
\centering
\includegraphics[width=0.72\textwidth]{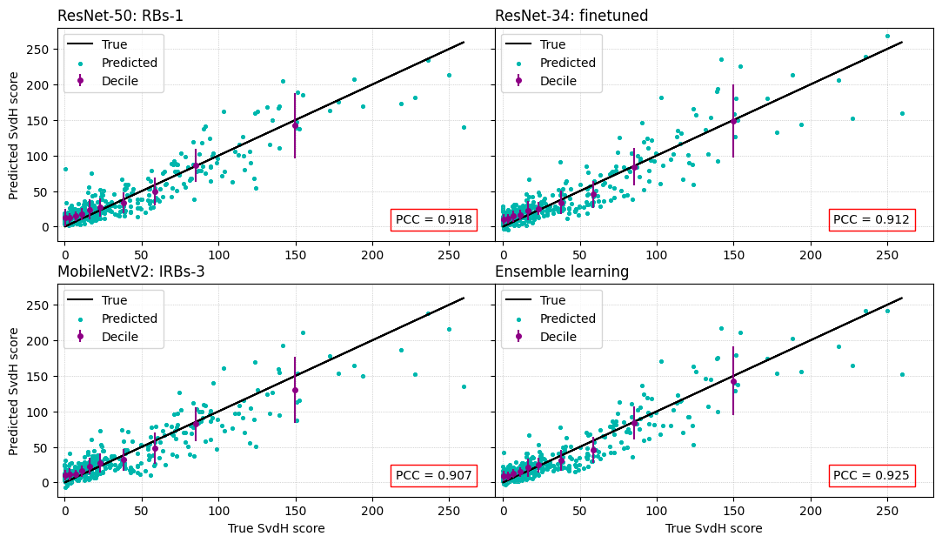}
\caption{Predicted vs. true SvdH scores for the best tuned regression models with transfer learning and the ensemble learning model that combines them. The mean and SD of the deciles are plotted for reference. The models yielded better performance in predicting early-stage cases. Ensemble learning reduced prediction errors on average compared to independent models.} \label{fig2}
\end{figure}

Transfer learning improved the performance of all three models irrespective of the tuning method (Table~\ref{tab1}). Among all regression models, ResNet-50:RBs-1 achieved the best overall accuracy, with a PCC of 0.918, MAE of 13.58 and RMSE of 18.88. 
For ResNet-34, finetuning increased the PCC by 3.6\% and decreased the MAE and RMSE by 1.27 and 2.92. 
MobileNetV2 yielded worse results than ResNets, and IRBs-3 only raised the PCC by 0.4\% and reduced the MAE and RMSE by 1.4 and 0.55. 
As shown in Fig.~\ref{fig2}, models' accuracy was better among cases with low SvdH scores. Their predictive error increased with the true score and several severe cases were underscored.

\subsubsection{Ensemble learning.}
Ensemble regressors combining ResNet-50:RBs-1, ResNet-34:finetuned and MobileNetV2:IRBs-3 were trained with different batch sizes. As shown in Table~\ref{tab1} and Fig.~\ref{fig2}, with batch size = 16, it outperformed all independent models, attaining a PCC of 0.925, MAE of 12.57 and RMSE of 18.02.

\subsubsection{Activation mapping.}
\begin{figure}[!ht]
\centering
\includegraphics[width=0.77\textwidth]
{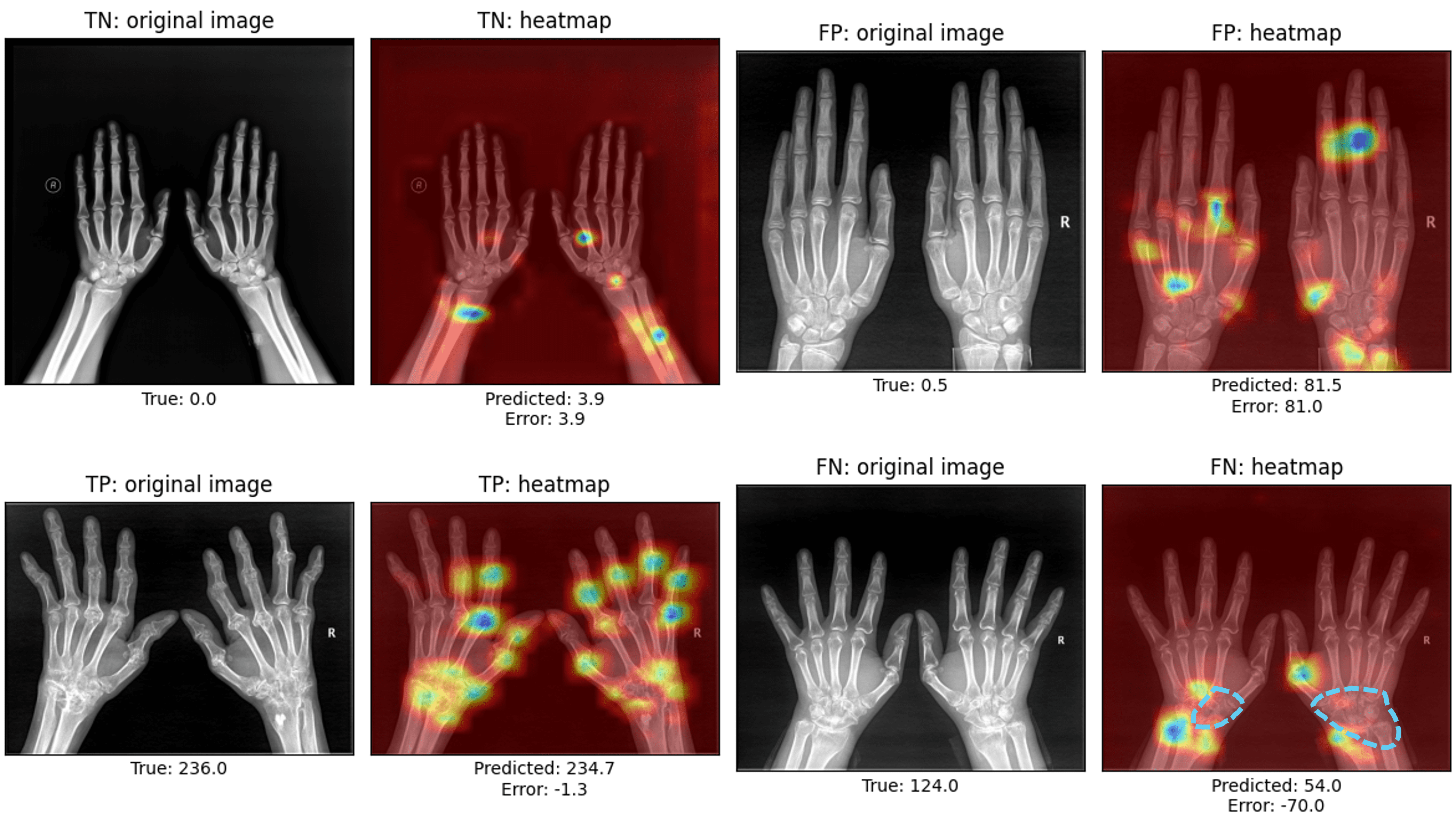}
\caption{Grad-CAM heatmaps of ResNet-50:RBs-1 regression model for examples of TN, FP, TP and FN. The true and predicted scores are provided. In FN, the overlooked changes in wrists are circled.} \label{fig3}
\end{figure}

The Grad-CAM heatmaps of TP, TN, FP, and FN examples predicted by the best independent regression model (i.e., ResNet-50:RBs-1) were provided in Fig.~\ref{fig3}. Except for TN, joints and bones assessed by the SvdH score were highlighted, suggesting that the model was looking at the desirable regions when making the prediction. However, in FP the highlighted joints did not demonstrate any pathology and the model failed to pick up a lot of sclerosis in the wrists in FN (depicted by circles in the heatmap). In TN, the diaphysis of some bones was highlighted.

\subsection{Classification task: RA severity class prediction}
\subsubsection{Baseline.}
\begin{figure}[!ht]
\centering
\includegraphics[width=\textwidth]{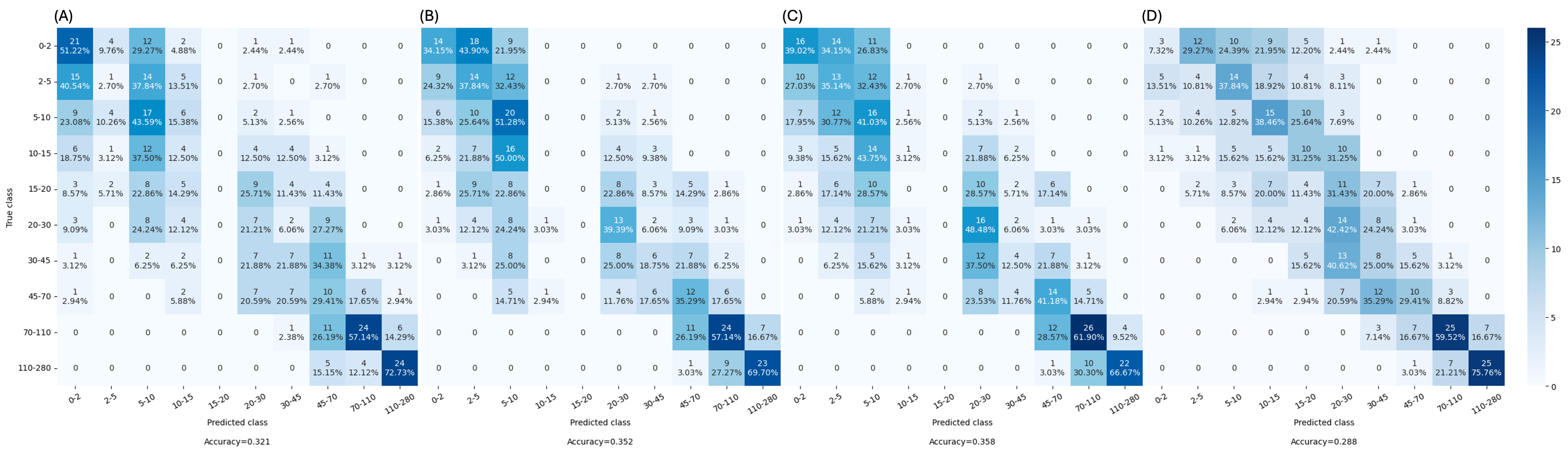}
\caption{Confusion matrices of (A) the best baseline classifier (ResNet-34), (B) the best pretrained classifier (ResNet-50:finetuned), (C) the best ensemble classifier, and (D) the ensemble regression model. The numbers and proportions of images from the same class that fell into different predicted classes are provided. The ensemble classifier yielded the smallest misclassifications and the highest overall accuracy.} \label{fig4}
\end{figure}

As shown in Table~\ref{tab2}, in general using a batch size of 4 yielded better results in terms of accuracy and BA. However, when referring to MAE, RMSE and PCC, MobileNetV2 trained with a batch size of 16 performed better. ResNet-50 achieved the highest accuracy (0.341) and BA (0.327). MobileNetV2 achieved the highest PCC (0.833). ResNet-34 achieved the lowest MAE (1.24) and RMSE (1.76). Balancing all four metrics, ResNet-34 demonstrated the best performance. Its confusion matrix is provided in Fig.~\ref{fig4}(A). Although the accuracy was low, especially for class 15-20, most mistakes were made between neighbouring classes and the misclassification rate decreased with severity.

\begin{table}[!ht]
\centering
\caption{Model performance in RA severity class prediction. The top two values of a metric are made bold and underlined. ResNet-34, ResNet-50:finetuned, and the ensemble classifier that combines class-specific predictions were the best-performing baseline, transfer learning and ensemble learning models. Substantial performance improvement was observed with transfer and ensemble learning. Using the best regression model for severity classification yielded the highest PCC and RMSE but much lower accuracy. }\label{tab2}
\resizebox{0.83\textwidth}{!}{\begin{tabular}{llllllll}
\multicolumn{8}{l}{\cellcolor[HTML]{F8CCA6}\textbf{Baseline}}                                                                                                                                                                                                                                                                                                                                                                          \\ \hline
\multicolumn{1}{|l|}{Model}                                                                                                                        & \multicolumn{2}{l|}{Batch size}                                                       & \multicolumn{1}{l|}{Accuracy}       & \multicolumn{1}{l|}{BA}             & \multicolumn{1}{l|}{PCC}            & \multicolumn{1}{l|}{MAE}           & \multicolumn{1}{l|}{RMSE}          \\ \hline
\multicolumn{1}{|l|}{}                                                                                                                             & \multicolumn{2}{l|}{4}                                                                & \multicolumn{1}{l|}{\textbf{0.341}} & \multicolumn{1}{l|}{\textbf{0.327}} & \multicolumn{1}{l|}{0.809}          & \multicolumn{1}{l|}{1.42}          & \multicolumn{1}{l|}{2.09}          \\ \cline{2-8} 
\multicolumn{1}{|l|}{\multirow{-2}{*}{ResNet-50}}                                                                                                  & \multicolumn{2}{l|}{16}                                                               & \multicolumn{1}{l|}{0.249}          & \multicolumn{1}{l|}{0.242}          & \multicolumn{1}{l|}{0.75}           & \multicolumn{1}{l|}{1.65}          & \multicolumn{1}{l|}{2.31}          \\ \hline
\multicolumn{1}{|l|}{}                                                                                                                             & \multicolumn{2}{l|}{4}                                                                & \multicolumn{1}{l|}{{\ul{0.321}}}    & \multicolumn{1}{l|}{{\ul{0.312}}}    & \multicolumn{1}{l|}{{\ul{0.829}}}    & \multicolumn{1}{l|}{\textbf{1.24}} & \multicolumn{1}{l|}{\textbf{1.76}} \\ \cline{2-8} 
\multicolumn{1}{|l|}{\multirow{-2}{*}{ResNet-34}}                                                                                                  & \multicolumn{2}{l|}{16}                                                               & \multicolumn{1}{l|}{0.296}          & \multicolumn{1}{l|}{0.289}          & \multicolumn{1}{l|}{0.784}          & \multicolumn{1}{l|}{1.36}          & \multicolumn{1}{l|}{1.92}          \\ \hline
\multicolumn{1}{|l|}{}                                                                                                                             & \multicolumn{2}{l|}{4}                                                                & \multicolumn{1}{l|}{0.316}          & \multicolumn{1}{l|}{0.31}           & \multicolumn{1}{l|}{0.822}          & \multicolumn{1}{l|}{1.33}          & \multicolumn{1}{l|}{1.86}          \\ \cline{2-8} 
\multicolumn{1}{|l|}{\multirow{-2}{*}{MobileNet V2}}                                                                                               & \multicolumn{2}{l|}{16}                                                               & \multicolumn{1}{l|}{0.313}          & \multicolumn{1}{l|}{0.301}          & \multicolumn{1}{l|}{\textbf{0.833}} & \multicolumn{1}{l|}{{\ul{1.24}}}    & \multicolumn{1}{l|}{{\ul{1.77}}}    \\ \hline
\multicolumn{8}{l}{\cellcolor[HTML]{B5EAF9}\textbf{Transfer learning}}                                                                                                                                                                                                                                                                                                                                                                 \\ \hline
\multicolumn{1}{|l|}{Model}                                                                                                                        & \multicolumn{1}{l|}{Layer freezing}                 & \multicolumn{1}{l|}{Batch size} & \multicolumn{1}{l|}{Accuracy}       & \multicolumn{1}{l|}{BA}             & \multicolumn{1}{l|}{PCC}            & \multicolumn{1}{l|}{MAE}           & \multicolumn{1}{l|}{RMSE}          \\ \hline
\multicolumn{1}{|l|}{}                                                                                                                             & \multicolumn{1}{l|}{No}                             & \multicolumn{1}{l|}{4}          & \multicolumn{1}{l|}{{\ul{0.352}}}    & \multicolumn{1}{l|}{{\ul{0.344}}}    & \multicolumn{1}{l|}{{\ul{0.842}}}    & \multicolumn{1}{l|}{{\ul{1.18}}}    & \multicolumn{1}{l|}{1.72}          \\ \cline{2-8} 
\multicolumn{1}{|l|}{}                                                                                                                             & \multicolumn{1}{l|}{RBs-1}                          & \multicolumn{1}{l|}{4}          & \multicolumn{1}{l|}{0.335}          & \multicolumn{1}{l|}{0.326}          & \multicolumn{1}{l|}{0.833}          & \multicolumn{1}{l|}{1.23}          & \multicolumn{1}{l|}{1.75}          \\ \cline{2-8} 
\multicolumn{1}{|l|}{\multirow{-3}{*}{ResNet-50}}                                                                                                  & \multicolumn{1}{l|}{RBs-2}                          & \multicolumn{1}{l|}{4}          & \multicolumn{1}{l|}{0.316}          & \multicolumn{1}{l|}{0.302}          & \multicolumn{1}{l|}{0.82}           & \multicolumn{1}{l|}{1.29}          & \multicolumn{1}{l|}{1.84}          \\ \hline
\multicolumn{1}{|l|}{}                                                                                                                             & \multicolumn{1}{l|}{No}                             & \multicolumn{1}{l|}{4}          & \multicolumn{1}{l|}{0.318}          & \multicolumn{1}{l|}{0.307}          & \multicolumn{1}{l|}{\textbf{0.844}} & \multicolumn{1}{l|}{1.2}           & \multicolumn{1}{l|}{{\ul{1.67}}}    \\ \cline{2-8} 
\multicolumn{1}{|l|}{}                                                                                                                             & \multicolumn{1}{l|}{RBs-1}                          & \multicolumn{1}{l|}{4}          & \multicolumn{1}{l|}{0.296}          & \multicolumn{1}{l|}{0.288}          & \multicolumn{1}{l|}{0.828}          & \multicolumn{1}{l|}{1.27}          & \multicolumn{1}{l|}{1.81}          \\ \cline{2-8} 
\multicolumn{1}{|l|}{\multirow{-3}{*}{ResNet-34}}                                                                                                  & \multicolumn{1}{l|}{RBs-2}                          & \multicolumn{1}{l|}{4}          & \multicolumn{1}{l|}{\textbf{0.358}} & \multicolumn{1}{l|}{\textbf{0.351}} & \multicolumn{1}{l|}{0.834}          & \multicolumn{1}{l|}{1.22}          & \multicolumn{1}{l|}{1.8}           \\ \hline
\multicolumn{1}{|l|}{}                                                                                                                             & \multicolumn{1}{l|}{No}                             & \multicolumn{1}{l|}{4}          & \multicolumn{1}{l|}{0.349}          & \multicolumn{1}{l|}{0.337}          & \multicolumn{1}{l|}{0.834}          & \multicolumn{1}{l|}{1.23}          & \multicolumn{1}{l|}{1.73}          \\ \cline{2-8} 
\multicolumn{1}{|l|}{}                                                                                                                             & \multicolumn{1}{l|}{IRBs-2}                         & \multicolumn{1}{l|}{4}          & \multicolumn{1}{l|}{0.341}          & \multicolumn{1}{l|}{0.327}          & \multicolumn{1}{l|}{0.819}          & \multicolumn{1}{l|}{1.24}          & \multicolumn{1}{l|}{1.79}          \\ \cline{2-8} 
\multicolumn{1}{|l|}{\multirow{-3}{*}{MobileNetV2}}                                                                                                & \multicolumn{1}{l|}{IRBs-3}                         & \multicolumn{1}{l|}{4}          & \multicolumn{1}{l|}{0.349}          & \multicolumn{1}{l|}{0.339}          & \multicolumn{1}{l|}{0.832}          & \multicolumn{1}{l|}{\textbf{1.17}} & \multicolumn{1}{l|}{\textbf{1.67}} \\ \hline
\multicolumn{8}{l}{\cellcolor[HTML]{BFFF99}\textbf{Ensemble learning}}                                                                                                                                                                                                                                                                                                                                                                 \\ \hline
\multicolumn{1}{|l|}{Model combination}                                                                                                            & \multicolumn{1}{l|}{Stacking principle}             & \multicolumn{1}{l|}{Batch size} & \multicolumn{1}{l|}{Accuracy}       & \multicolumn{1}{l|}{BA}             & \multicolumn{1}{l|}{PCC}            & \multicolumn{1}{l|}{MAE}           & \multicolumn{1}{l|}{RMSE}          \\ \hline
\multicolumn{1}{|l|}{}                                                                                                                             & \multicolumn{1}{l|}{}                               & \multicolumn{1}{l|}{4}          & \multicolumn{1}{l|}{{\ul{0.355}}}    & \multicolumn{1}{l|}{{\ul{0.348}}}    & \multicolumn{1}{l|}{{\ul{0.854}}}    & \multicolumn{1}{l|}{{\ul{1.13}}}    & \multicolumn{1}{l|}{1.64}          \\ \cline{3-8} 
\multicolumn{1}{|l|}{}                                                                                                                             & \multicolumn{1}{l|}{\multirow{-2}{*}{All classes}}  & \multicolumn{1}{l|}{16}         & \multicolumn{1}{l|}{0.355}          & \multicolumn{1}{l|}{0.342}          & \multicolumn{1}{l|}{0.851}          & \multicolumn{1}{l|}{1.19}          & \multicolumn{1}{l|}{1.71}          \\ \cline{2-8} 
\multicolumn{1}{|l|}{}                                                                                                                             & \multicolumn{1}{l|}{}                               & \multicolumn{1}{l|}{4}          & \multicolumn{1}{l|}{0.352}          & \multicolumn{1}{l|}{0.344}          & \multicolumn{1}{l|}{0.853}          & \multicolumn{1}{l|}{{\ul{1.13}}}    & \multicolumn{1}{l|}{{\ul{1.63}}}    \\ \cline{3-8} 
\multicolumn{1}{|l|}{\multirow{-4}{*}{\begin{tabular}[c]{@{}l@{}}ResNet-50:finetuned \& \\ ResNet-34:RBs-2 \& \\ MobileNetV2:IRBs-3\end{tabular}}} & \multicolumn{1}{l|}{\multirow{-2}{*}{Single class}} & \multicolumn{1}{l|}{16}         & \multicolumn{1}{l|}{\textbf{0.358}} & \multicolumn{1}{l|}{\textbf{0.349}} & \multicolumn{1}{l|}{\textbf{0.859}} & \multicolumn{1}{l|}{\textbf{1.12}} & \multicolumn{1}{l|}{\textbf{1.62}} \\ \hline
\multicolumn{8}{l}{\cellcolor[HTML]{FADDDC}\textbf{Regression to classification}}                                                                                                                                                                                                                                                                                                                                                      \\ \hline
\multicolumn{2}{|l|}{Ensemble regression model}                                                                                                                                                          & \multicolumn{1}{l|}{16}         & \multicolumn{1}{l|}{0.288}          & \multicolumn{1}{l|}{0.29}           & \multicolumn{1}{l|}{\textbf{0.861}} & \multicolumn{1}{l|}{1.14}          & \multicolumn{1}{l|}{\textbf{1.54}} \\ \hline
\end{tabular}
}
\end{table}

\subsubsection{Transfer learning.}
As shown in Table~\ref{tab2}, model pretraining improved the performance of MobileNetV2 irrespective of the tuning scheme, with IRBs-3 achieving the highest accuracy (0.349) and BA (0.339) and lowest MAE (1.17) and RMSE (1.67). Transfer learning’s effect on ResNets was less consistent. Though improvements in the regression metrics were observed for all tuning schemes of ResNet-50, only finetuning improved the classification accuracy (0.352) and BA (0.344), with a PCC of 0.842, MAE of 1.18, and RMSE of 1.72. For ResNet-34, though finetuning yielded the best PCC, MAE and RMSE, RBs-2 was the only scheme attaining a higher accuracy (0.358) and BA (0.351) than the baseline. 
Balancing all metrics, we concluded that ResNet-50:finetuned was the best model. As shown in Fig.~\ref{fig4}(B), there were fewer misclassifications between far-away classes, which is reflected by the rise in PCC and drop in MAE and RMSE.

\subsubsection{Ensemble learning.}
Stacking class-specific outputs of ResNet-50:finetuned, ResNet-34:RBs-2, and MobileNetV2:IRBs-3 to predict the probability of a single class yielded the best results with batch size = 16 (see Table~\ref{tab2}). Achieving a PCC of 0.859, MAE of 1.12 and RMSE of 1.62, it outperformed all independent classifiers.

\subsubsection{Activation mapping.}

\begin{figure}[!ht]
\centering
\includegraphics[width=0.77\textwidth]{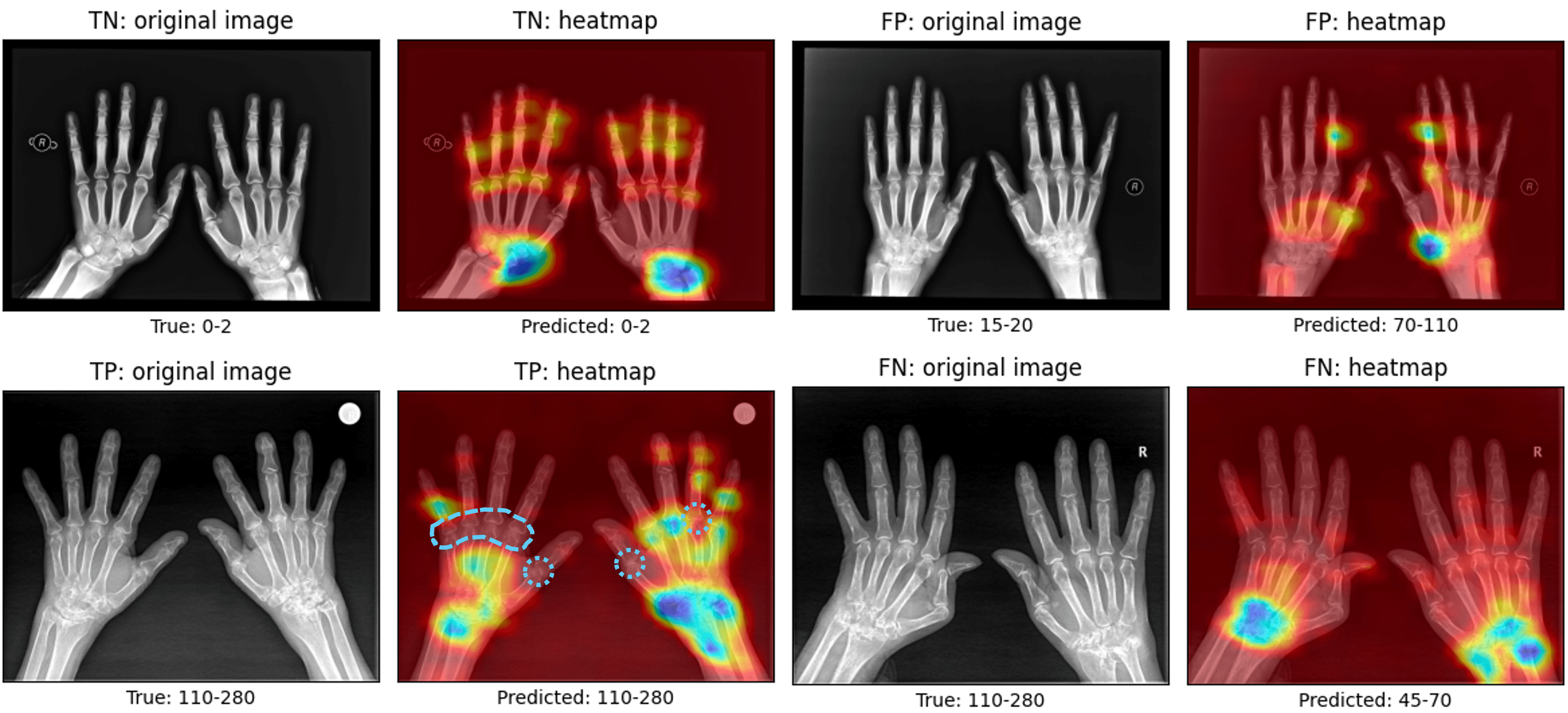}
\caption{Grad-CAM heatmaps of ResNet-50:finetuned classifier for examples of TN, FP, TP and FN. In TP, some overlooked changes in fingers are circled.} \label{fig5}
\end{figure}

Fig.~\ref{fig5} displays the Grad-CAM heatmaps extracted from the best independent classifier (ResNet-50:finetuned) for TP, TN, FP and FN examples. Despite highlighting the joints and bones assessed in SvdH scoring, the heatmaps show relatively high activation at regions of irrelevant anatomical structures, such as the diaphysis of metacarpals and the distal interphalangeal joints in TP. Although the damages in the wrists were picked up in TP, several finger bones and joints in the left hand and the metacarpophalangeal joints in the right hand were overlooked (circled in the heatmap). 
In TN, normal joints in the wrists were highlighted. The reason for the high intensity around the right thumb is unclear, but this could be an area of confusion for the model because of a high prevalence of osteoarthritis in this area.
Despite looking at the correct area (only the wrists as the fingers are normal) for damage, the model underestimated the RA severity of FN. In FP, the identified right thumb base displays some irregularities that do not map to the score, thus might confuse our model.

\subsubsection{Regression to classification.}
As shown in Table~\ref{tab2} and Fig.~\ref{fig4}(D), the ensemble regression model achieved higher PCC (0.861) and RMSE (1.54) than the best ensemble classifier in severity classification but much lower accuracy (0.288) and BA (0.29), especially among samples with SvdH $<$ 10. Its performance was better than the classification model in classes 10-15 and 15-20.

\section{Discussion}
Our results on SvdH score prediction suggested that MSE was a more suitable loss function than the smooth loss proposed in \cite{Wang2022}. Without pretraining, MobileNetV2 outperformed the ResNets, which contradicts Wang et al.’s observations. When trained from scratch, the baseline models achieved better results than ResNet-Dwise50 in the paper, which obtained a PCC of 0.95, MAE of 17.76, and RMSE of 26.34 without transfer learning. Our baseline MobileNetV2’s testing MAE and RMSE were 2.77 and 5.86 lower. In this work, ResNet-34 outperformed ResNet-50, suggesting that the improvement might result from a more lightweight model architecture that reduces the risk of overfitting. 

Among all baseline classifiers, ResNet-34 achieved the best performance. Considering the high MAE and RMSE between the scores assigned by two radiologists, we expected there would be more errors when predicting milder cases as narrower classes were set. Since cross-entropy loss cannot account for the ordinal relationships between classes, our models with higher classification accuracy did not always demonstrate better performance in terms of regression metrics. Other loss functions which penalise bigger errors could be adopted in future studies.

Our results with transfer learning showed the potential of optimising RA scoring models’ performance by pretraining them in a larger dataset of similar types of images. Although the RSNA challenge images were from children and single-hand, the models still acquired useful features for our tasks. This is additionally supported by the improved performance when tuning models with early layers frozen. Unlike models trained from scratch for score prediction, tuned ResNet-50 achieved the best performance while tuned MobileNetV2 performed the worst. One potential reason might be that ResNet-50 could learn more features from the bigger pretraining dataset that later became useful for SvdH score prediction. Also, it might suffer less from overfitting in the larger dataset. With transfer learning, more complex model architectures could be experimented with. ResNet-Dwise50 pretrained in CheXpert achieved a PCC of 0.97, MAE of 14.9 and RMSE of 22.01. Our best independent regression model (ResNet-50:RBs-1) outperformed it by further reducing the MAE by 1.32 and RMSE by 3.13. Furthermore, our model achieved almost as good RMSE as the radiologists and its MAE was only 1.34 higher. However, the PCC was much lower, suggesting scope for further improvement. As shown by the activation maps, the model could identify and focus on the joints in RA images. However, it might not pick up all the pathologies as shown by FN and might confuse healthy joints with damaged joints as in FP. In future, other medical imaging datasets of similar or different types of anatomical structures such as RadImageNet \cite{Mei2022} could be adopted for transfer learning to acquire more comprehensive feature representations.
The performance of the regression models deteriorated in more severe cases, probably because of the reduction in sample size. In future model optimisation, methods to handle data imbalance should be introduced. 

For classification, no specific patterns in the results of different model tuning schemes were observed and some set-ups even worsened the prediction accuracy despite improving the other metrics. This again shows the weaknesses of cross-entropy loss and emphasises the need for a more suitable loss for ordinal classification problems. The activation maps suggested that the model also looked at some anatomical structures that are irrelevant to SvdH scoring. Since late-stage patients may fail to straighten their fingers during an X-ray, it is important to investigate if the positioning of fingers influences the prediction of both regression and classification models. Our Grad-CAM results did not suggest this issue, but additional testing is needed to improve the trustworthiness of the models.

Ensemble learning of the best pretrained independent models yielded further performance improvement in both regression and classification tasks. The ensemble regressor even achieved smaller RMSE than radiologists. Stacking has effectively combined the strengths of each model and evened out the failures. 
Though not directly comparable, regression models outperformed classification models in terms of PCC and provided higher resolution in the predictions.
When using the best regression model for classification, the higher prediction resolution resulted in higher PCC and lower RMSE. Since the entire score range was treated equally during regression training, it performed much worse than the best classifier in classifying very early-stage cases.

\section{Conclusion}
Our results of predicting the SvdH score and SvdH-based severity class of hand radiographs using state-of-the-art CNN models demonstrated the great potential of DL techniques in quantifying RA on a whole image level. Transfer learning using a paediatric single-hand X-ray dataset and ensemble learning improved the prediction accuracy substantially, resulting in models that largely outperformed previously proposed models \cite{Wang2022} and achieved performance close to the experienced radiologists. For score prediction, future work should explore ways to deal with data imbalance to reduce large errors in severe cases and improve the correlation between predictions and true scores. For classification, alternative loss functions and model architectures could be considered to improve prediction accuracy and reduce distances of misclassification.

\subsubsection{Acknowledgements.}
Zhiyan Bo is supported by the EPSRC Centre for Doctoral Training in Health Data Science (EP/S02428X/1). We would like to thank the Oxford Biomedical Research Computing Facility for providing the computational resources.

\bibliographystyle{splncs04}
\bibliography{DLforRAscoring_240613}

\end{document}

\title{Supplementary Material}
\titlerunning{Deep Learning Models to Score Hand Radiographs for Rheumatoid Arthritis}

\author{Zhiyan Bo\inst{1}\textsuperscript{(\Letter)}\orcidlink{0009-0002-6458-3156}\and
Laura C. Coates\inst{2}\orcidlink{0000-0002-4756-663X}
\and Bart\l omiej W. Papie\.z\inst{1}\orcidlink{0000-0002-8432-2511}
}
\authorrunning{Z. Bo et al.}

\institute{Big Data Institute, Nuffield Department of Population Health, University of Oxford, Oxford, UK \\
\email{zhiyan.bo@reuben.ox.ac.uk} \&
\email{bartlomiej.papiez@bdi.ox.ac.uk}\\
\and
Nuffield Department of Orthopaedics, Rheumatology and Musculoskeletal Sciences, University of Oxford, Oxford, UK}

\maketitle              

\section{The van der Heijde modification of the Sharp score}

The van der Heijde-modified Sharp (SvdH) score \cite{Heijde1992} assesses bone erosion in 16 areas in each hand and wrist. These include four proximal interphalangeal (PIP) joints, interphalangeal (IP) joint, five metacarpophalangeal (MCP) joints, the first metacarpal, multangulars (trapezium and trapezoid), scaphoid, lunate, radius, and ulna. The erosion score is cumulative, with:
\begin{itemize}
  \item 0 = normal;
  \item 1 = discrete erosion;
  \item 2 = large erosion not passing midline;
  \item 3 = large erosion passing midline.
\end{itemize}
The maximum erosion score per area in the hands is 5, which could be a sum of multiple erosions. For example, a joint with two small discrete erosions scores 2 ($1+1$), while a joint with two large erosions not passing the midline and 2 small erosions scores 5 ($2+2+1+1 = 6 \to 5$). 
Therefore the total erosion score accounting for both hands and wrists ranges from 0 to 160 ($2 \times 16 \times 5$).

The SvdH score assesses joint space narrowing (JSN) in 15 joints in each hand and wrist, which include four PIP, five MCP, the third, fourth, and fifth carpometacarpal (CMC), trapezium-scaphoid, capitate-scaphoid-lunate, and radiocarpal joints. The JSN score ranges from 0 to 4, with:
\begin{itemize}
  \item 0 = normal;
  \item 1 = focal or doubtful;
  \item 2 = general, $\le$ 50\% of the original joint space;
  \item 3 = general, $>$ 50\% of the original joint space, or subluxation;
  \item 4 = ankylosis.
\end{itemize}
Therefore the total JSN score accounting for both hands and wrists ranges from 0 to 120 ($2 \times 15 \times 4$).

Combining the bone erosion and JSN scores, the total SvdH score for hands and wrists ranges from 0 to 280. A low score means little RA-related damage, while a high score means severe damage and/or reduced mobility of multiple joints. 

\bibliographystyle{splncs04}
\bibliography{supplementary_material_240613}